\documentclass[12pt,a4paper]{article}
\usepackage[preprint]{dfttrob}
\dfttnum{DFTT 73/98\\FISIST/2-99/CENTRA\\February 15, 1999}

\def\Tevatron{\textsc{Tevatron}}

\def\nbar{\bar n}            
\def\Nbar{\bar N}            
\def\nc{{\bar n_c}}          

\def\pt{p\kern -.2pt\lower 4pt\hbox{\tiny T}}    

\def\p0{P_0(\Delta y)}

\def\NF{\mathcal{N}_{\kern -1.9pt f}}
\def\NC{\mathcal{N}_{\kern -1.7pt c}}

\usepackage{amsmath}
\usepackage{amssymb}
\usepackage{epsfig}

\def\ksoft{k_{\text{soft}}}
\def\nsoft{\nbar_{\text{soft}}}
\def\Nsoft{\Nbar_{\text{soft}}}

\def\ksemi{k_{\text{semi-hard}}}
\def\nsemi{\nbar_{\text{semi-hard}}}
\def\Nsemi{\Nbar_{\text{semi-hard}}}

\def\asoft{\alpha_{\text{soft}}}
\def\ktot{k_{\text{total}}}
\def\ntot{\nbar_{\text{total}}}

\def\rs{\sqrt{s}}

\begin{document}

\title{Possible scenarios for soft and semi-hard components  structure
            in central hadron--hadron collisions in the TeV region:\\ 
            pseudo-rapidity intervals.}
\author{A. Giovannini\\
 \it Dipartimento di Fisica Teorica and
   I.N.F.N -- sezione di Torino\\
 \it via P. Giuria 1, 10124 Torino, Italy\\[3mm]
 R. Ugoccioni\\
 \it CENTRA and Departamento de F{\'\i}sica (I.S.T.),\\ 
 \it Av. Rovisco Pais, 1096 Lisboa codex, Portugal}
\maketitle

\begin{abstract}
Continuing previous work on collective variables properties in full phase 
space in hadron-hadron collisions in the TeV region our investigation is 
extended to multiplicity distributions and clan structure analysis in
pseudo-rapidity intervals. Total multiplicity distributions (MDs) 
are considered also here as the weighted superposition of soft and semi-hard 
components, and described by Pascal(NB)MD. 
  The soft component is characterised by KNO and 
Feynman scaling behaviour in all scenarios which differ  in the semi-hard 
component properties only. In fact, the semi-hard component 
has been supposed to satisfy,  in scenario 1, KNO and Feynman scaling
behaviour 
and to violate it  strongly in scenario 2. A third possibility has been 
explored: it is a QCD inspired scenario and leads to expectations 
intermediate between the just mentioned two. 
  The semi-hard component structure becomes dominant in the TeV energy domain,
a huge mini-jet production is indeed  the main phenomenon in this region.
  In addition a new 
species of clans is generated suggesting a phase transition in the clan 
production in scenarios 2 and 3. Our results can be compared  with 
\Tevatron\ and  LHC  experimental data 
when available.       
\end{abstract}

\section{Introduction}
In the first paper of this series \cite{combo:prd}
(from now on referred to as `I'), possible scenarios for
collective variables properties in the TeV region have been examined
in full phase space.
  Stated that shoulder structure in $P_n$ vs.~$n$ and $H_q$ vs.~$q$
oscillations can be interpreted for c.m.\ energies larger than 200 GeV
as the effect of the weighted
superposition of soft and semi-hard events, each class being described
by a single Pascal (also known as negative binomial \cite{combo:prd}) 
multiplicity distribution, (Pa(NB)MD) 
\begin{equation}
   P^{\text{(PaNB)}}_n(\nbar,k) = \frac{k(k+1)\dots(k+n-1)}{n!}
			\frac{\nbar^n k^k}{(\nbar+k)^{n+k}}
\end{equation}
our approach consisted in finding physically motivated
extrapolations of the free parameters of the
mentioned distributions starting from their known behaviour
in the GeV energy region.
  This fact has led us to imagine firstly possible extreme scenarios in
the TeV region which basically should fix upper and lower bounds to
the allowed variation path of the average multiplicity $\nbar_i$ and
of the parameter $k_i$, which is linked to the dispersion $D_i$ by
\begin{equation}
  (D_i^2 - \nbar_i)/\nbar_i^2 = 1/k_i			\label{eq:disp}
\end{equation}
  Here $i$ stands for `soft' or `semi-hard'. 
  Accordingly, in scenario 1 we assumed KNO scaling to be valid both for
$P_{n,\text{soft}}$ and $P_{n,\text{semi-hard}}$ multiplicity
distributions (MDs); consequently, $\ksoft$ and $\ksemi$ parameters
were taken constant with energy.
  In scenario 2, KNO scaling is realized for the soft
component only ($\ksoft$ constant with energy) and $1/\ksemi$
increases linearly in $\ln(\rs)$.
  Between these two quite extreme possibilities we proposed 
a third scenario, which in view of the chosen behaviour of
the parameters 
of $P_{n,\text{semi-hard}}$ was called a QCD inspired scenario
(the scenarios are described in greater detail 
in Section \ref{sec:scenarios}).

It is interesting to remark that data on MDs at 1.8 TeV c.m.\ energy
(from the E735 experiment \cite{Walker}), when compared with our
predictions, are closer to  scenario 2,
characterised by a huge mini-jet production in the semi-hard component,
but go beyond it showing an even wider MD: assuming these data
will be confirmed, observed deviations from expectations of scenario 2
might very well indicate the onset in our framework of new
substructures in the total MDs, which we suggested to interpret as
probably due to a new species of mini-jets (see I).
Our caution on this point was and is motivated by the fact that
mentioned data on MDs in full phase space (f.p.s.) at lower c.m.\
energies show systematic differences with respect to UA5 data
\cite{UA5:rep} on which is based our general scheme for defining
scenarios 1 and 2. That's the reason why we decided to maintain this
scheme also for extending our previous work from f.p.s. to
pseudo-rapidity intervals. When data of the E735 experiment will be
consolidated it is indeed not a too hard job to adapt our approach to
the new experimental framework.

\section{$P_n$ vs $n$ behaviour in pseudo-rapidity intervals
in the GeV and TeV energy domains}
In going from full phase space (f.p.s.) to pseudo-rapidity ($\eta$) intervals,
our main concern is to be consistent with the scenarios explored in
f.p.s., and extend them.

In f.p.s., the quadratic growth (in $\ln \rs$) of the total average
multiplicity was attributed to the growing contribution of semi-hard
events. 
Notice that semi-hard events are defined
by the presence of mini-jets or jets in the final state,
irrespectively of the pseudo-rapidity interval under consideration.
It must be stressed that only after
this classification of events has been carried out we look at phase
space intervals: thus the description of the total MD,
$P_n(\eta_c,\rs)$, in terms of
a weighted superposition of two multiplicity distributions holds in
$\eta$ intervals with the same weighting factor as in f.p.s., 
namely $\asoft$, function of energy only and not of $\eta_c$
(the $\eta_c$ dependence comes from $\nbar$ and $k$ parameters), i.e.:
\begin{equation}	\label{eq:combo.eta}
\begin{split}
P_n(\eta_c,\rs) = &
  \asoft(\rs) P_n^{\text{(PaNB)}}\left(\nsoft(\eta_c,\rs),
  \ksoft(\eta_c,\rs)\right) + \\
     &\left(1 -\asoft(\rs)\right) P_n^{\text{(PaNB)}}\left(\nsemi(\eta_c,\rs),
     \ksemi(\eta_c,\rs)\right)
\end{split}
\end{equation}
precisely as in eq.~(I.2).

In this paper, we will be concerned with symmetric 
pseudo-rapidity intervals
$[-\eta_c, \eta_c]$, with $1 \le \eta_c \le 3$. The joining of these
intervals to f.p.s.\ is assumed to be smooth.

\subsection{Average multiplicity in pseudo-rapidity intervals}
In f.p.s. (see I, Section 2.1), it was assumed that each 
component has an average
multiplicity which grows linearly with $\ln \rs$:
\begin{gather*}
\nsoft(\rs) = -5.54 +4.72 \ln(\rs) \tag{I.3} \\
\nsemi(\rs) \approx 2 \nsoft(\rs)  \tag{I.4.A}
\end{gather*}

Since the width
of available phase space also grows linearly with $\ln \rs$, we find
that the simplest way to be consistent with our assumptions is to
say that the single particle density must show an energy independent
plateau around $\eta = 0$ which extends some units in each
direction 
(a plateau of this size is found in experimental data at UA5
energies for the full distribution: its height increases with
c.m.\ energy indicating violation of Feynman scaling.)

Numerically, we fix the height $\nbar_0$ of the soft and semi-hard plateaus
again respecting the result of \cite{Fug} in the investigation
of UA5 data:
\begin{equation}
  \nbar_{0,\text{soft}} \approx 2.45 , \qquad\qquad
      \nbar_{0,\text{semi-hard}} \approx 6.4
\end{equation}
and
\begin{equation}
  \nbar_i(\eta_c) = 2 \nbar_{0,i} \eta_c \qquad(i =
      \text{soft,semi-hard})
\end{equation}
Accordingly, from eq.\ \eqref{eq:combo.eta}
\begin{equation}
  \nbar_{\text{total}}(\eta_c,\rs) = \asoft(\rs)
      \nbar_{\text{soft}}(\eta_c) +
	\left( 1 - \asoft(\rs)\right) \nbar_{\text{semi-hard}}(\eta_c)
  \label{eq:combo.eta.n}
\end{equation}
where the last line follows from eq.\ \eqref{eq:combo.eta}. Notice
that the semi-hard component is more than twice the soft component,
and the value 2.45 for the soft component is compatible with low
energy data (e.g., ISR data), where only the soft component is
present.

There are no compelling physical reasons to assume that also the
semi-hard component has an energy independent plateau. Indeed a
logarithmic growth of the plateau with c.m.\ energy is compatible with
a second possibility that was considered in I for the growth of
$\nsemi$:
\begin{equation*}
\nsemi(\rs) \approx 2 \nsoft(\rs)   + 0.1 \ln^2(\rs) \tag{I.4.B}
\end{equation*} 
In the simplest approach where one
neglects energy variations of $d \nbar/d \eta$ at the boundary of phase
space, a parameterisation of the growth numerically compatible
with eq.~(I.4.B) is
\begin{equation}
  \nbar_{0,\text{semi-hard}} \approx 6.3 + 0.07 \ln\rs
\end{equation}
the effect of which in the 1--20 TeV range is in complete agreement
with f.p.s. Therefore we limit ourselves to showing figures only for
the case of linear $\nsoft$,
mentioning the differences in the text below when relevant.
We postpone to future work the discussion of the case in which
the particle density varies at the boundary of phase space.

\subsection{Dispersion in pseudo-rapidity intervals}
We now examine the width of the multiplicity distribution; to this
end, we use the parameter $k$ as defined in eq.~\eqref{eq:disp}.
In particular, we have the following relation:
\begin{equation}
  \ntot^2 \left(1+\frac{1}{\ktot}\right) = \asoft
      \nsoft^2 \left(1+\frac{1}{\ksoft}\right) +
	\left( 1 - \asoft \right) \nsemi^2
	       \left(1+\frac{1}{\ksemi}\right)
  \label{eq:combo.eta.k}
\end{equation}
obtained from eq.s \eqref{eq:combo.eta} and \eqref{eq:combo.eta.n} (for brevity,
the dependence on $\eta_c$ and $\rs$ has been omitted in this
formula). 
The behaviour of $\ksoft$ and $\ksemi$ is indeed of great importance
in our subsequent discussion for at least three reasons.
Firstly, in view of $k$'s relationship with the two-particle
correlation function $C_2(\eta_1,\eta_2;\rs)$, \cite{AGLVH:1}:
\begin{equation}
  k^{-1}(\eta_c;\rs) = \frac{1}{\nbar^2(\eta_c;\rs)} \iint_{-\eta_c}^{\eta_c} 
	C_2 (\eta_1,\eta_2;\rs) \, d\eta_1 d\eta_2
\end{equation}
$\ksoft$ and $\ksemi$
control two-particle correlation properties of the two components.
Secondly, clan structure parameters, $\Nbar_i(\eta_c,\rs)$ and
$\nbar_{c,i}(\eta_c,\rs)$ ($i$ = soft, semi-hard),
are defined for each component
in terms of $\nbar_i$ and $k_i$ as follows \cite{AGLVH:1}:
\begin{gather}
  \Nbar_i(\eta_c,\rs) = k_i(\eta_c,\rs) 
				\ln\left( 1+\nbar_i(\eta_c)/k_i(\eta_c,\rs) \right) ; \notag \\
  \nbar_{c,i}(\eta_c,\rs) = \nbar_i(\eta_c) / \Nbar_i(\eta_c,\rs)
                   \label{eq:clandef}
\end{gather}
It should be pointed out that 
the above definition is valid for a single Pa(NB)MD only:
as explained in \cite{combo:prd}, clans cannot be defined
for the total MD which, being the superposition of two (or
possibly more) Pa(NB)MDs with in general different parameters,
is not of Pa(NB)MD type.
The third reason is a consequence of eq.~\eqref{eq:clandef}, which
allows us to interpret $1/k_i$ for a single Pa(NB)MD only:
\begin{equation}
 k_i^{-1} = \frac{\mathtt{P}_i(1;2)}{\mathtt{P}_i(2;2)}
\end{equation}
where $\mathtt{P}_i(N;m)$ is the probability to have $m$ particles belonging
to $N$ clans \cite{AGLVH:4}. Therefore any assumption or result
on the energy or pseudo-rapidity dependence of $k_i$ has its
counterpart in all above mentioned frameworks.

\begin{table}
\caption[Values of 1/k]{Values of the parameters assumed in our extrapolations
for $1/k = (D^2-\nbar)/\nbar^2$, for each component, for each rapidity
interval examined and for f.p.s., too.
For the soft one and for
scenario 1, $1/k$ is energy independent and given in the table; in the
other cases the relevant parameters are given. 
}\label{tab:extraps}
\begin{center}
\begin{tabular}{@{\extracolsep{12pt}}ccccc}
\hline
interval  & soft comp. & scenario 1 & scenario 2 & scenario 3 \\[2mm]
$|\eta| \le \eta_c$ & $\ksoft^{-1}$ & $\ksemi^{-1}$ &   
   $k_{\text{total}}^{-1}(\eta_c,\rs) =$ &
   $k_{\text{semi-hard}}^{-1}(\eta_c,\rs) =$ \\[1mm]
 &&&  $a + b\ln\rs$ & 
   $C + D/{\sqrt{\ln(\rs/10)}}$\\[3mm]
\hline
$\eta_c = 1$ & 0.294 & 0.217 & 
	\parbox[c][1.5\height][c]{2.5cm}{$a = 0.02$\\$b = 0.08$} & 
	\parbox{3cm}{$C = 0.97$ \\$D = -1.6$} \\
\hline
$\eta_c = 2$ & 0.286 & 0.172 & 
	\parbox[c][1.5\height][c]{2.5cm}{$a = -0.06$\\$b = 0.08$} & 
	\parbox{3cm}{$C = 0.88$ \\$D = -1.5$} \\
\hline
$\eta_c = 3$ & 0.250 & 0.156 & 
	\parbox[c][1.5\height][c]{2.5cm}{$a = -0.12$\\$b = 0.08$} & 
	\parbox{3cm}{$C = 0.72$ \\$D = -1.2$} \\
\hline
f.p.s &        0.143 & 0.077 & 
	\parbox[c][1.5\height][c]{2.5cm}{$a = -0.082$\\$b =0.0512$} & 
	\parbox{3cm}{$C = 0.38$ \\$D = -0.42$}\\
\hline
\end{tabular}
\end{center}
\end{table}

\begin{figure}
\begin{center}\mbox{\epsfig{file=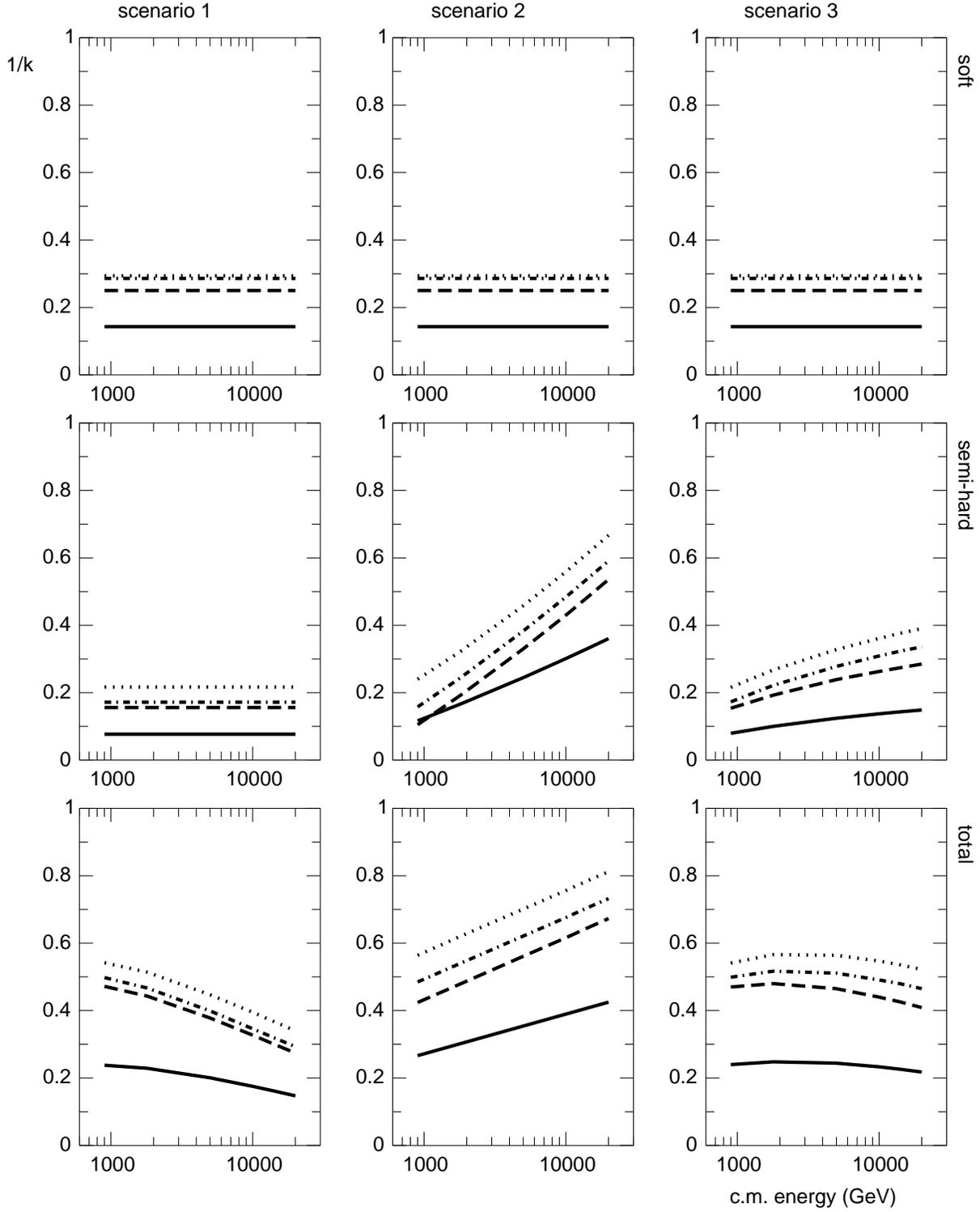,width=16cm}}
\end{center}
\caption[1 over k in rapidity intervals]{The Pa(NB)MD parameter $1/k$
is plotted against the c.m.\ energy for three rapidity intervals
(dotted line: $\eta_c=1$; dash-dotted line: $\eta_c=2$;
dashed line: $\eta_c=3$)
and for f.p.s. (solid line), for each scenario (in columns, from left
to right: scenario 1, 2 and 3) and for each component (in rows, from top to
bottom: soft, semi-hard, total distribution).}\label{fig:k.sdy}
\end{figure}

The soft component is taken to have
$1/k$ constant with energy for each $\eta$ interval,
but variable with the width of the interval. In low energy
experimental data, $1/k$ is not constant but KNO scaling holds:
in view of the growing value of $\nbar$, KNO scaling implies at high
energies that $1/k$ reaches a constant value, which we infer from
the highest energy data point in reference \cite{Fug}. For the actual
numbers, see table \ref{tab:extraps}; 
the behaviour of all the relevant Pa(NB)MD parameters
is shown in Figures~\ref{fig:k.sdy}, \ref{fig:clan.sdy} and
\ref{fig:nc.sdy}.
We notice that $1/k$ decreases slowly by increasing the width of the
$\eta$ interval: as the interval gets larger, there is 
less aggregation. Particles generated by new clans fill the growing
interval faster than those generated by old clans. Accordingly the linear
growth of the average number of clans $\Nbar$ is faster than the
increase in the average number of particles per clan, $\nc$. 
This behaviour also implies that, having clans a large extension 
in pseudo-rapidity, long range correlations become important.

In summary, for the soft component both $\nbar$ and $k$ are constant
with energy: so are the clan parameters. $\nbar$, $\Nbar$ and $\nc$
all grow with $\eta_c$, while $1/k$ decreases.

\section{The three scenarios in pseudo-rapidity intervals}\label{sec:scenarios}
For the semi-hard component, since
in f.p.s.\ we devised three scenarios with a different variation of the
dispersion of the multiplicity distribution with energy, 
we use the fact that low energy experimental results for the 
dispersion show the same energy behaviour in $\eta$ intervals as in
f.p.s., and extend the f.p.s.\ behaviour to pseudo-rapidity intervals.

\subsection{Scenario 1}
The first scenario is characterized by a Feynman scaling and
KNO scaling semi-hard component, just as for the soft component, but
with different values for the parameters, see table \ref{tab:extraps}
and the leftmost column of Figure~\ref{fig:k.sdy}:
the average multiplicity is more than double, and $1/k$ is
smaller, so that even with more particles there is less aggregation.
In this case, the total distribution's $\ktot$ parameter is given by
the superposition formula, Eq.\ \eqref{eq:combo.eta.k}.

It is interesting to notice in connection with this scenario that
correlations increase due to the superposition of events of different
type, both with smaller correlations, as
\begin{equation}
  1/\ktot > 1/\ksoft > 1/\ksemi
\end{equation}
This is an example of the situation examined in detail in 
\cite{correl}: these enhanced correlations are
a consequence of the fluctuations
in single particle densities (due to the superposition of events with
different average multiplicity),
superimposed to ``genuine'' two-particle correlations.

Furthermore, the behaviour of $1/\ktot$ with energy is peculiar in
that it first increases (up to about 1 TeV) then decreases: the
maximum is rather wide, resulting in an accidental KNO scaling
behaviour for c.m.\ energy $0.5 \lesssim \rs \lesssim 1.8$ TeV.

The KNO scaling behaviour of the total MD is unexpected because
although we are superimposing two KNO scaling distributions, we are
doing it with an energy dependent weight parameter. Of course scaling
behaviour is expected both at low energy (only the soft component is
present) and at very high energy, because in this simple picture only
the semi-hard component is present.

\begin{figure}
\begin{center}\mbox{\epsfig{file=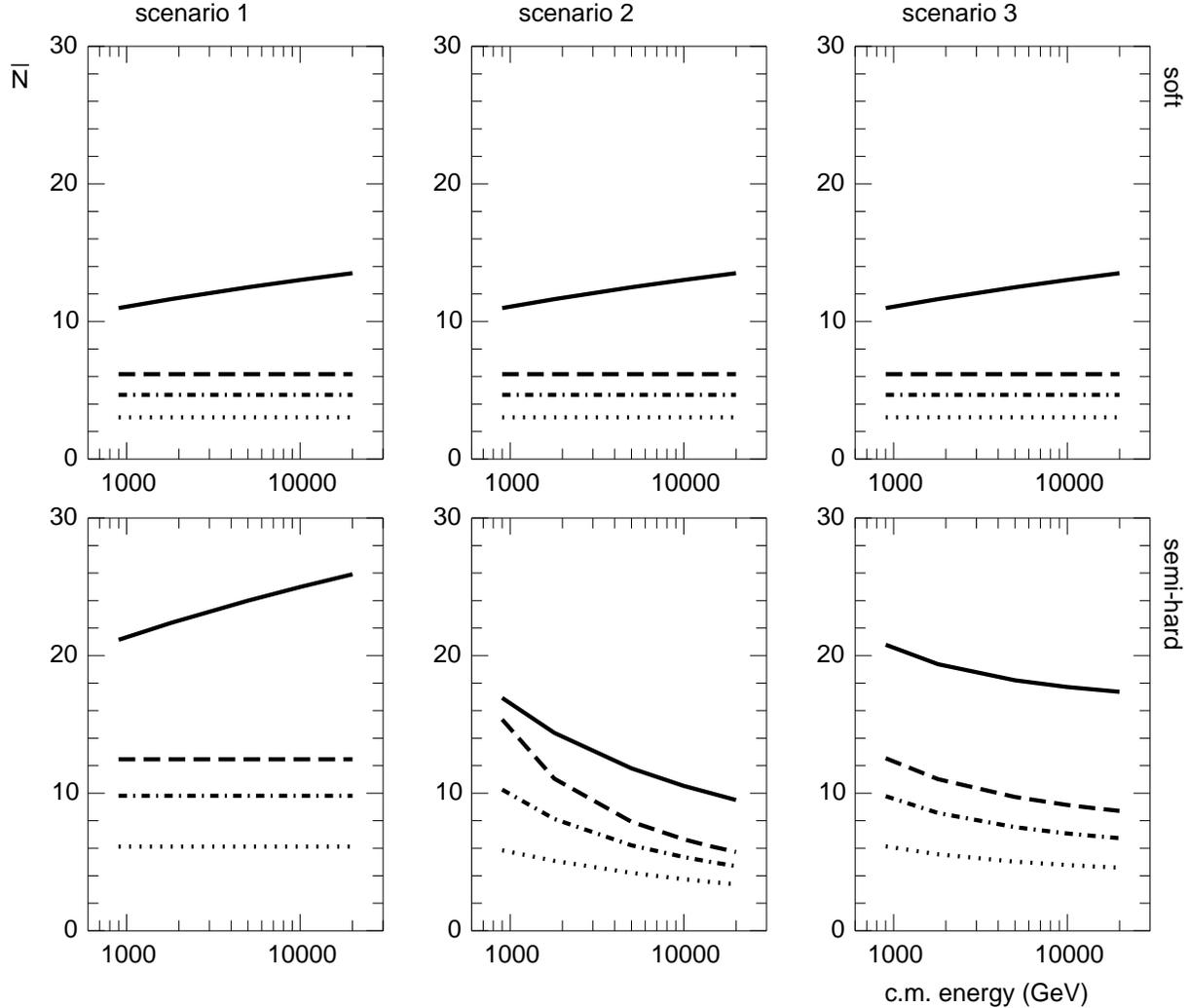,width=16cm}}
\end{center}
\caption[Nclan in rapidity intervals]{The
average number of clans $\bar N$
is plotted against the c.m.\ energy for three rapidity intervals
(dotted line: $\eta_c=1$; dash-dotted line: $\eta_c=2$;
dashed line: $\eta_c=3$)
and for f.p.s. (solid line), for each scenario (in columns, from left
to right: scenario 1, 2 and 3) and for each component (in rows, from top to
bottom: soft and semi-hard).}\label{fig:clan.sdy}
\end{figure}

\begin{figure}
\begin{center}\mbox{\epsfig{file=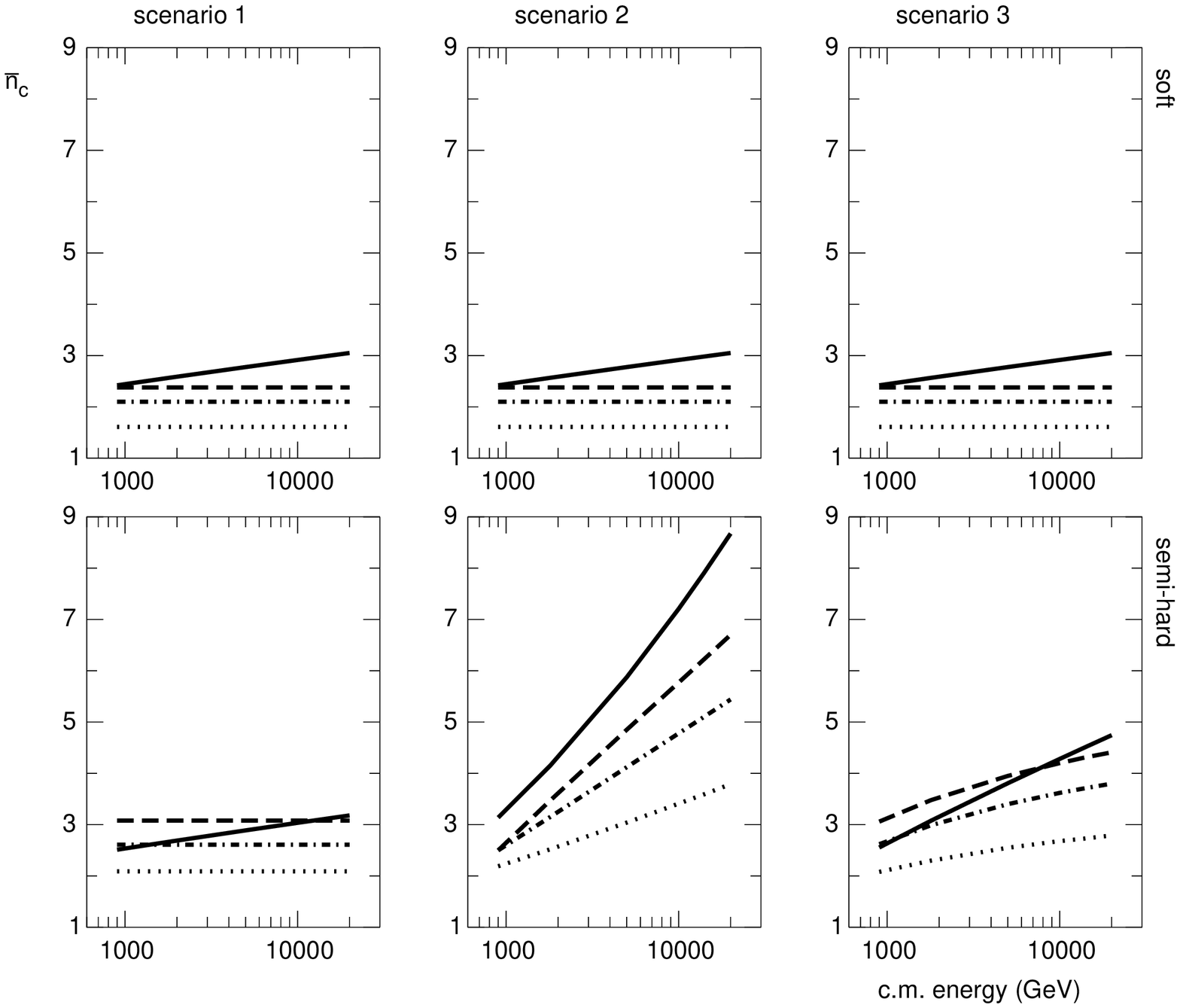,width=16cm}}
\end{center}
\caption[nc in rapidity intervals]{The 
average number of particles per clan, $\nc$
is plotted against the c.m.\ energy for three rapidity intervals
(dotted line: $\eta_c=1$; dash-dotted line: $\eta_c=2$;
dashed line: $\eta_c=3$)
and for f.p.s. (solid line), for each scenario (in columns, from left
to right: scenario 1, 2 and 3) and for each component (in rows, from top to
bottom: soft and semi-hard).}\label{fig:nc.sdy}
\end{figure}

The energy independence of $\nbar$ and $k$ in 
fixed $\eta$ intervals is contrasted
with their energy dependence in f.p.s. in terms of clan parameters
in the leftmost column in Figures \ref{fig:clan.sdy} and \ref{fig:nc.sdy}.

The effect of a quadratic growth of $\nsemi$ with energy is to
increase slightly (around 10\%) the value of $1/\ktot$, compatibly
with what we have seen in f.p.s.

\subsection{Scenario 2}
For the second scenario we choose to violate KNO scaling by making
$1/\ktot$ continue to grow with energy as it does up to UA5 energies
with a linear behaviour in $\ln \rs$ as given in table
\ref{tab:extraps}, where the parameters $a$ and $b$ have been fitted
to experimental data from ISR to UA5.
Notice that it appears that the best slope $b$ is the same for each
interval. $1/\ksemi$ is then obtained using eq.\ \eqref{eq:combo.eta.k}:
it also grows approximately linearly with c.m.\ energy, and
decreases rapidly as $\eta_c$ increases
(see Figure~\ref{fig:k.sdy}, central column).
In particular, above 1 TeV,
$1/\ksemi$ becomes larger than $1/\ksoft$: 
this implies that correlations are much larger
in the semi-hard events than in the soft events; because in both cases
$k < \nbar$, this is probably due again to fluctuations in 
single particle distribution (the semi-hard component has indeed larger
fluctuations in multiplicity). 

At the same time, the average number of clans is seen to decrease very
rapidly with the energy for the semi-hard component, $\nc$ is seen
to increase with energy; it also increases with $\eta_c$, and the
increase is faster when the energy is  higher
(Figures \ref{fig:clan.sdy} and \ref{fig:nc.sdy}.)

The effect of a quadratic growth of $\nsemi$ with energy is to
decrease slightly (less than 10\%) the value of $1/\ksemi$, compatibly
with what we have seen in f.p.s.

\subsection{Scenario 3}
In the third scenario we chose a QCD inspired shape, which has
a behaviour which turns out to be intermediate between scenario 1 and 2:
it starts growing with energy
but asymptotically (well above the energy range we consider here)
tends to a constant value:
\begin{equation}
  \frac{1}{\ksemi} = C + \frac{D}{\sqrt{\ln(\rs/10)}}
	\label{eq:k.3}
\end{equation}
Again the values of the parameters for each interval
are given in table \ref{tab:extraps}: they were chosen to be
compatible with the 900 GeV points \cite{Fug} and to lead to an
intermediate value of $1/k$.
The general behaviour of the Pa(NB)MD parameters is shown in
the rightmost column of 
Figures~\ref{fig:k.sdy}, \ref{fig:clan.sdy} and \ref{fig:nc.sdy}.

While the behaviour of the parameters for the semi-hard component
is qualitatively similar to that of scenario 2, the behaviour for the
total MD is qualitatively similar to that of scenario 1.
Indeed, the increase of $1/k$ for the semi-hard with c.m.\ energy is 
not as fast as in scenario 2, and $1/k$ is smaller in this case, 
so this leads, for the total distribution,
to a broad maximum in the energy range 2-10 TeV, which implies KNO
scaling. The decrease from the maximum is slower than in scenario 1,
and this accidental KNO scaling appears at higher energies.

The effect of a quadratic growth of $\nsemi$ with energy is to
increase slightly (around 10\%) the value of $1/\ktot$, compatibly
with what we have seen in f.p.s.

\section{Comments on the three scenarios}
It is quite clear that  in scenario 1 both soft and semi-hard components  show
wide self-similarity regions \cite{treleani}: 
the parameters $\ksoft$ and $\ksemi$ vary
very little  from one pseudo-rapidity interval to another. A quite strong
point, which can  easily be tested by using  Pa(NB)MD  with a fixed 
 $k$ (soft or semi-hard) parameter (determined by data in a small domain of
rapidity space)  as a ``microscope'':  by enlarging 
slowly the initial domain in rapidity one  can explore up to which interval 
MDs are described by Pa(NB)MDs with the same initial $k$. 
This exercise will tell us  that in that region two-particle correlations 
are dominant and that they vary according to the normalization 
$\nsemi^2(\eta_c)$ only.
  It should also be noticed that the fact that $k$ parameter is energy
independent in a fixed rapidity interval  and vary  very little from
one interval to another has important consequences on
$\Nsoft$ and $\Nsemi$ (see Fig.\ \ref{fig:clan.sdy}): 
they do not vary with energy 
in a fixed rapidity interval and only very slowly by increasing the
rapidity interval; their growth with energy
in full phase space is  is due to the growth of
the average number of particles with a constant $k$ parameter.

In scenario 2  the soft  component shows of course for all parameters the
same behaviour seen in other scenarios. The interest here is on the semi-hard
component structure and on its difference with that of scenario 1.
$\Nsemi$ in scenario 2 decreases very fast as the c.m.\  energy
increases, this trend  should be compared  with that of the semi-hard 
component in scenario 1: here $\Nsemi$ is an increasing function
of c.m.\  energy in f.p.s.\ and is constant in different 
pseudo-rapidity intervals.
Accordingly, $\nc$ (see Fig.\ \ref{fig:nc.sdy})
is growing very fast with c.m.\  energy in scenario 2; it is growing
very slowly in f.p.s. and is constant with energy in pseudo-rapidity 
intervals in scenario 1. These completely different clan structure behaviors  
 when KNO and Feynman scaling are satisfied (scenario 1) and violated 
(scenario 2) have an interesting interpretation.

Newly created particles of the semi-hard component in scenario 1, being 
their aggregation power ($1/\ksemi$) quite limited and energy independent,
give origin to clans  whose average number of particles is  an energy
independent quantity in rapidity intervals (very slowly growing with the 
 extension of the interval) and gently increasing  from $\approx$ 2.5 
at 1 TeV to $\approx$ 3 at 15 TeV in f.p.s.
In scenario 2 as the energy increases newly created particles not only 
continue to aggregate in the existing clans in view of the large value
of $1/\ksemi$  (if only this fact would
occur  $\Nsemi$  would be an energy independent quantity, a
situation which could  be true in scenario 2 for  the semi-hard component
only asymptotically) but in addition  $\Nsemi$ starts to decrease
in the TeV region, i.e., the aggregation is now involving clans themselves.
Clan aggregation into ``super-clans'' is an unexpected new phenomenon, which
occurs in all rapidity intervals and is less pronounced for 
pseudo-rapidity intervals of smaller size.
For energies much higher than
10 TeV  clan aggregation stops ($\Nsemi \approx$ constant) and the
new created particles continue to go in the existing super-clans.
Scenario 3  confirms for the semi-hard component its main peculiarity to 
have intermediate properties among those  of the semi-hard components
in scenarios 1 and 2.

\begin{figure}
\begin{center}
  \mbox{\epsfig{file=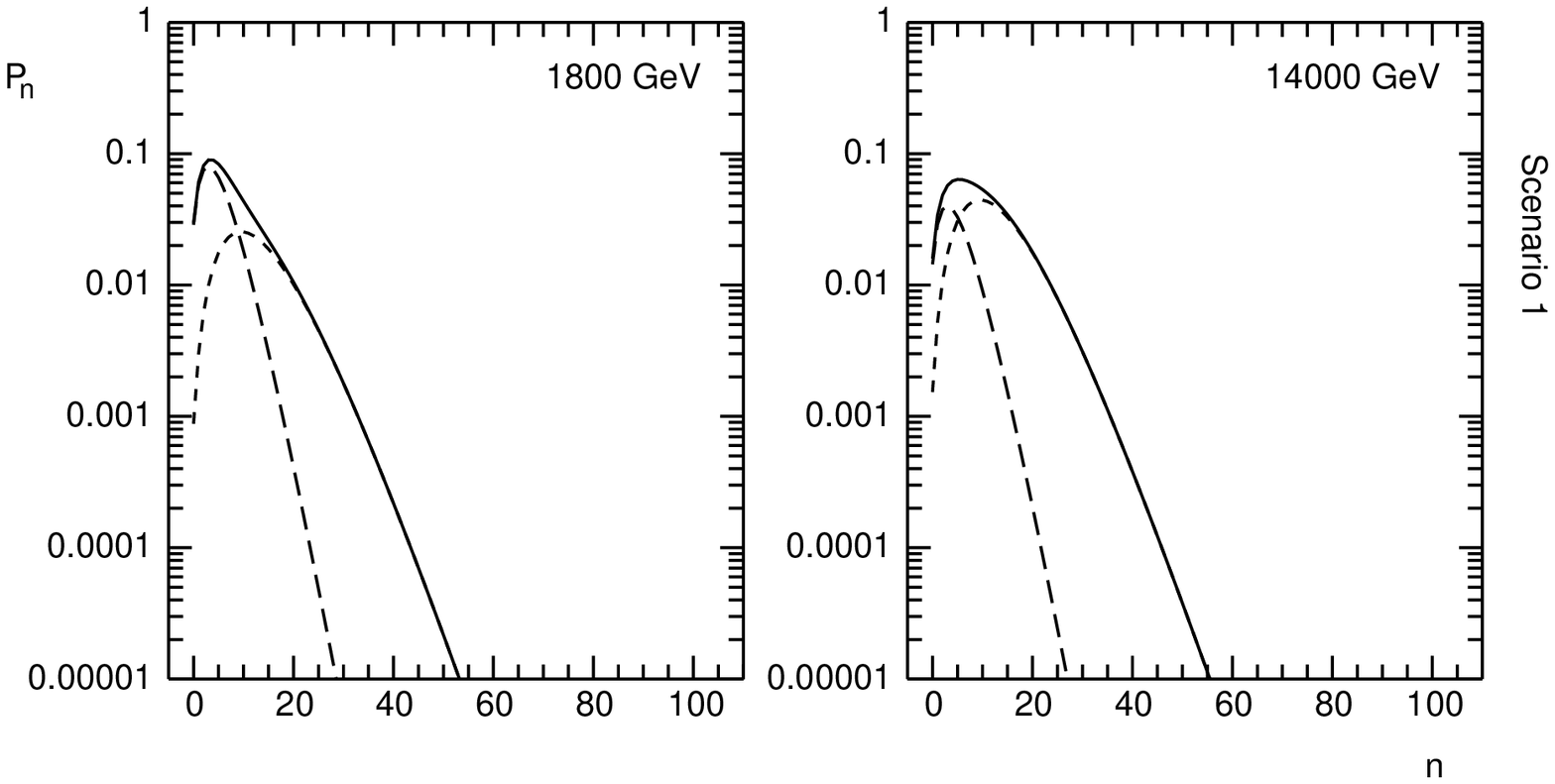,width=12cm}}
  \mbox{\epsfig{file=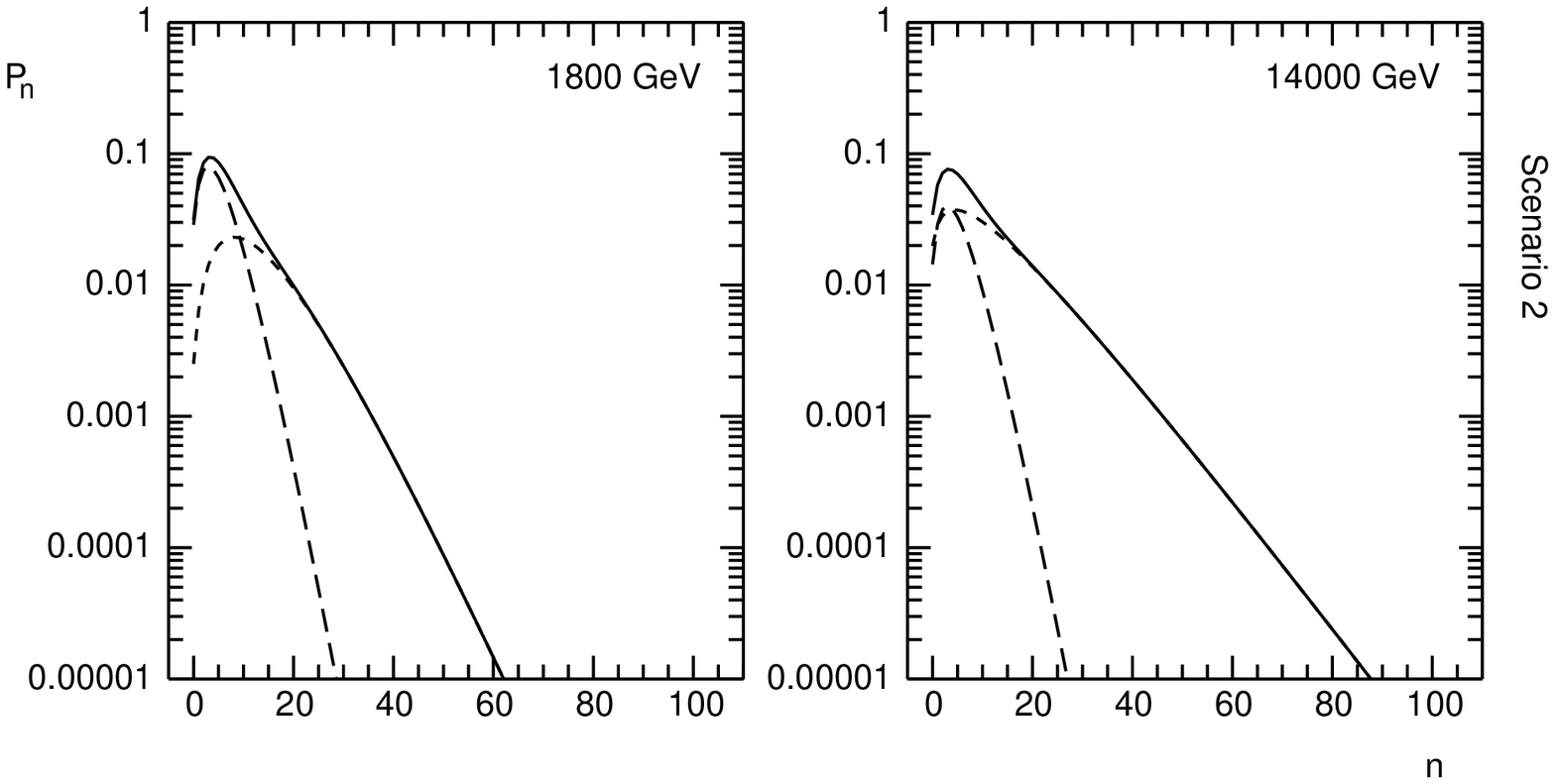,width=12cm}}
  \mbox{\epsfig{file=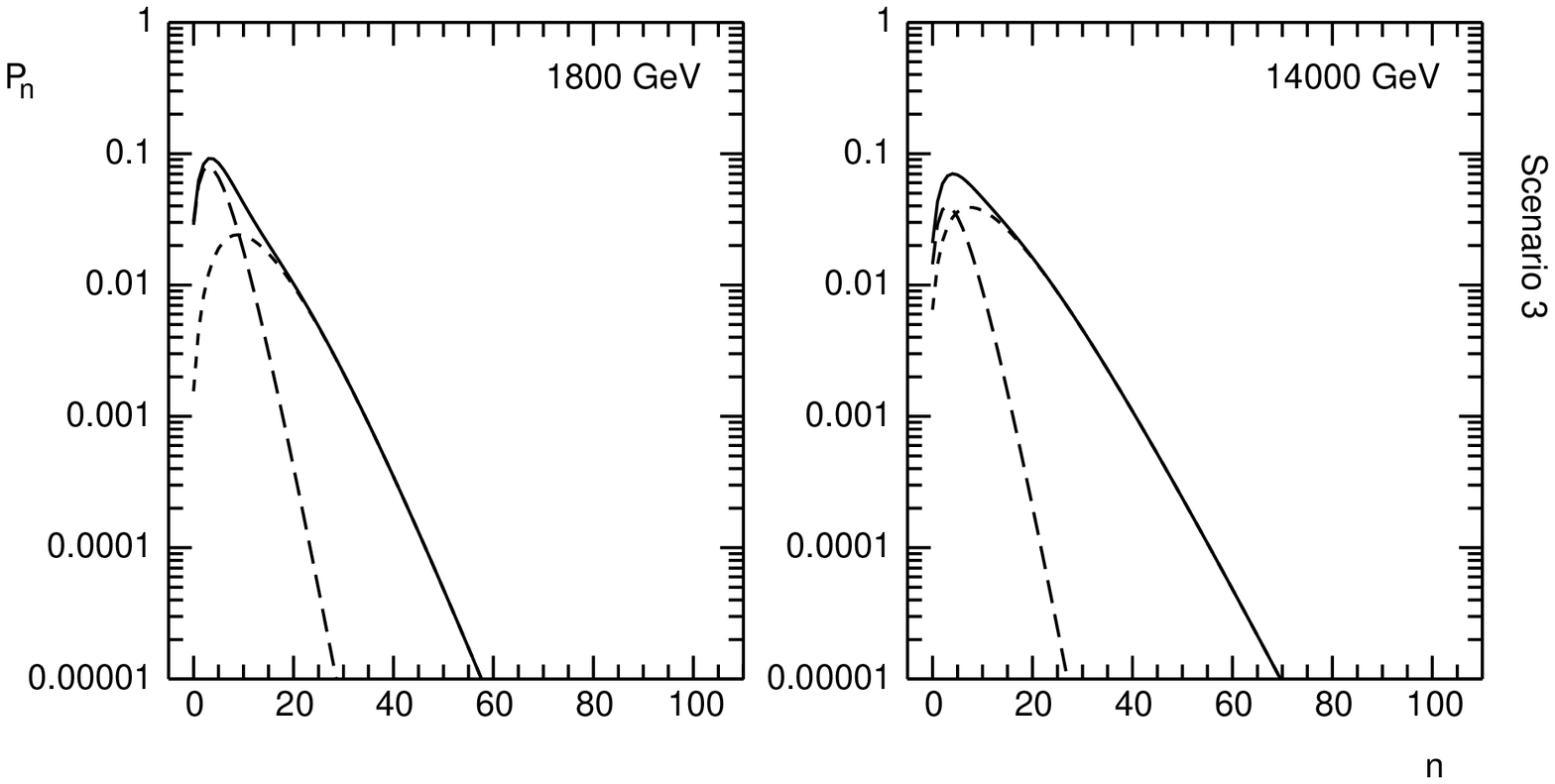,width=12cm}}
\end{center}
\caption[MD in intervals]{Multiplicity distributions 
for the pseudo-rapidity interval
$|\eta| < 1$, for the three scenarios (top to bottom: scenarios 1, 2,
3) at the c.m.\ energies of \Tevatron\ (1.8 TeV)
and LHC (14 TeV). Solid line: total distribution; dashed
line: soft component; sort-dashed line: semi-hard component.}\label{fig:md.yc1}
\end{figure}

In order to complete our study in Figure~\ref{fig:md.yc1}  
we show a comparison for the
multiplicity distributions  for the small interval $\eta_c  = 1$  at c.m.\ 
energy 1.8 TeV and  c.m.\  energy 14 TeV. (The figure should be compared
also with that one   shown in paper I  in full phase space).

We notice here how
the semi-hard component becomes dominant at 14 TeV c.m.\ energy,
although in the low multiplicity part of the distributions the soft
component is almost as large as the semi-hard one. These graphs
confirm the behaviour already seen in f.p.s.

From the analysis of these figures, we conclude that the interesting 
phenomenon  which clan structure analysis allowed to see
is not rapidly expanding.
This fact points out that in order to distinguish the different scenarios,
one should  look at $1/k$ parameters  and related clan structure
analysis, in particular the different behavior  between scenarios 1 and 2 is 
striking. In principle  it should be measurable already at the \Tevatron.
Scenario 3 is different as its parameters due to the lack of
precise QCD calculations on the matter are more flexible and can be 
adjusted to get close to either one of the other scenarios.

\section{Summary}
We  studied possible scenarios for soft and semi-hard components structures
in central hadron-hadron collisions  in the TeV region in symmetric
pseudo-rapidity intervals. The paper is the  natural extension
of  previous work on hadron-hadron collisions in full phase space
and has a twofold motivation. Firstly, in order to understand the dynamics of
multiparticle production and related correlations it is important to study 
their behaviour in regions  where the role of conservation laws is negligible.
Secondly, future particle accelerators in the TeV energy domain  are expected 
not to be equipped  with full acceptance detectors; they will explore 
limited  sectors of full phase space only, unfortunately. Accordingly, 
the  contact of theoretical expectations with experiments in the next decade 
should be looked for in pseudo-rapidity intervals  and not in full phase space.
Since we wanted   to avoid complications due to the presence
of the dip around zero  value of pseudo-rapidity variable  we considered
intervals greater than one unit in rapidity, and in order to  be sure 
that the influence of conservation laws is small we fixed the upper bound 
to three unit in pseudo-rapidity variable  to our intervals on  both sides
of the origin.

Selected intervals   are therefore wide enough (they extend up to
six units in rapidity) to allow significant predictions, and chosen  in   
regions not too small and far from the borders of phase space enough
to guarantee  results not affected  by the above mentioned problems.
Following  paper I and supported by data  at lower energies our
 main assumption   has been that  
soft and semi-hard events are  described by  Pascal (NB)  multiplicity 
distributions; in addition, the soft event fraction has been taken 
pseudo-rapidity (interval) independent and varying with center of mass energy 
only. Total  multiplicity distributions are the result of the weighted 
superposition of the two above mentioned  more elementary substructures.
Single particle  densities  develope in our picture an
energy independent central plateau  for
soft and semi-hard  components and their difference is limited to  the 
heights  of the corresponding two plateaus. The joining to full pase space 
is taken to be smooth for simplicity leaving more complex situations for 
future work.  

The soft component structure is fully characterized by a $\ksoft$  Pa(NB)MD 
parameter which is constant with energy  for each pseudo-rapidity interval but
varying with its width,  i.e., is characterized by Feynman 
and KNO scaling behaviour.

Three possible scenarios are discussed for the semi-hard component. 
Scenario 1 has Feynman and KNO scaling as for the soft component, but with
different values of the parameters ($1/\ksemi$ is energy independent
and varies very little with pseudo-rapidity intervals). In scenario 2
KNO scaling is violated and the  width of the total multiplicity distribution
grows linearly as $\ln\sqrt{s}$   ($1/\ksemi$ is quickly increasing
with energy and   decreasing with pseudo-rapidity intervals). In scenario 3
the  slope of $1/\ksemi$ is suggested by QCD: it grows initially with
 energy but asymptotically  (well above the  extreme values of the
abscissa allowed in  figure 1) it tends to a constant value; its
increase with energy is not as fast as in scenario 2, but its decrease
with pseudo-rapidity intervals quite similar.

In conclusion  $\ksoft$ show wide self-similarity regions   and the average 
number of (soft) clans in a fixed rapidity interval is an  energy independent 
quantity in all three scenarios. In scenario 1, $\ksemi$  behaves as 
$\ksoft$  but it has a larger value; being the average number of particles in 
the semi-hard sector larger than in the soft one,  clans of the semi-hard 
component are more numerous than clans of the soft component  and have a 
larger number of particles per clan.
In scenario 2 self-similarity appears only as an asymptotic property  for
the semi-hard component; the average number of clans is decreasing with energy
and as the pseudo-rapidity interval decreases but the average number of
particles per clan is becoming quite large as the energy increases.
Scenario 3 has predictions which are ---as expected--- intermediate 
between the previous two.

Of course now the word is to experiments. They will determine which
one of the discussed possibilities is closest to the real world. CDF
can help in this direction. It is a fact that one should expect the
dominance of the semi-hard component structure as the energy increases
also in pseudo rapidity intervals, i.e., huge mini-jets production is
also here the main characteristic in the new region, as shown
explicitly by MDs in a fixed pseudo-rapidity interval at different
energies in Figure~\ref{fig:md.yc1}.

In addition the semi-hard component behaviour has a suggestive
interpretation in all scenarios in terms of its clan properties.
In scenario 1 one notices numerous clan production of nearly equal size
as the energy increases in all rapidity intervals.
in scenario 2 in all pseudo-rapidity intervals  to the
aggregation of newly created particles into existing clans follows the
aggregation of clans themselves into super-clans (a new species of 
(mini)-jets) whose average number becomes at asymptotic energies  nearly 
constant. An interesting phenomenon resembling a phase transition in clan
production mechanism. Notice that the average number of clan is higher in
larger rapidity intervals and its decreasing with energy favours stronger
long range correlations.

Scenario 3 leads to predictions which are ---as usual--- 
intermediate between the
previous two extreme situations but in view of its flexibility can be 
modified in the two directions as long as QCD will not provide new constraints
on our formulae.

\section{Acknowledgements}
This work was supported in part by M.U.R.S.T. (under Grant 1997).
R.~U. would like to acknowledge the financial support of
the Portuguese Ministry of Science and Technology via the
``Sub-Programa Ci\^encia e Tecnologia do $2^o$ Quadro Comunit\'ario de
Apoio.''

\section*{References}

\end{document}